\documentclass[final,twocolumn,pre,aps,amsfonts,amssymb,showpacs,superscriptaddress,floatfix]{revtex4-1}
\usepackage{amsmath,mathtools}
\usepackage{graphicx}
\usepackage{color}
\usepackage{hyperref}
\usepackage{psfrag}
\usepackage{stmaryrd}
\usepackage{amssymb}
\usepackage{wasysym}
\usepackage[latin1]{inputenc}

\begin{document}

\title{Disentangling inertial waves from eddy turbulence in a forced rotating
turbulence experiment}

\author{Antoine Campagne}
\affiliation{Laboratoire FAST,
CNRS, Universit\'e Paris-Sud, Orsay, France}
\author{Basile Gallet}
\affiliation{Laboratoire SPHYNX, Service de
Physique de l'\'Etat Condens\'e, CEA Saclay, CNRS UMR 3680, 91191
Gif-sur-Yvette, France}
\author{Fr\'{e}d\'{e}ric Moisy}
\affiliation{Laboratoire FAST, CNRS,
Universit\'e Paris-Sud, Orsay, France}
\author{Pierre-Philippe Cortet}
\affiliation{Laboratoire FAST,
CNRS, Universit\'e Paris-Sud, Orsay, France}

\date{\today}

\begin{abstract}

We present a spatio-temporal analysis of a statistically
stationary rotating turbulence experiment, aiming to extract a
signature of inertial waves, and to determine the scales and
frequencies at which they can be detected. The analysis uses
two-point spatial correlations of the temporal Fourier transform
of velocity fields obtained from time-resolved stereoscopic
particle image velocimetry measurements in the rotating frame. We
quantify the degree of anisotropy of turbulence as a function of
frequency and spatial scale. We show that this
space-time-dependent anisotropy is well described by the
dispersion relation of linear inertial waves at large scale, while
smaller scales are dominated by the sweeping of the waves by fluid
motion at larger scales. This sweeping effect is mostly due to the
low-frequency quasi-two-dimensional component of the turbulent
flow, a prominent feature of our experiment which is not accounted
for by wave turbulence theory. These results question the
relevance of this theory for rotating turbulence at the moderate
Rossby numbers accessible in laboratory experiments, which are
relevant to most geophysical and astrophysical flows.

\end{abstract}

\maketitle

\section{Introduction}

The energy content of turbulence is usually characterized by the
energy distribution among spatial scales, either in physical or in
Fourier space. For rotating, stratified, or magnetohydrodynamic
turbulence~\cite{DavidsonBook2013}, waves can propagate and
coexist with ``classical'' eddies and coherent structures, which
advocates for a spatio-temporal description of such flows. While
temporal fluctuations are usually slaved to the spatial ones via
sweeping effects in classical
turbulence~\cite{Chen1989,Sanada1992}, they are expected to be
governed by the dispersion relation of the waves for time scales
much smaller than the eddy turnover time. The latter regime is the
subject of wave turbulence theory, in which the assumption of weak
nonlinear coupling between waves allows to predict scaling laws
for the spatial energy spectrum~\cite{Zakharov1992,Nazarenko2011}.

It is a matter of debate whether wave turbulence theory (also
known as weak turbulence theory) is a good candidate to describe
rotating turbulence in the rapidly rotating limit. Solutions to
the linearized rotating Euler equation can be decomposed into
inertial waves, which satisfy the anisotropic dispersion relation
\begin{equation}
\sigma({\bf k}) = 2\Omega \frac{|k_\parallel|}{|{\bf k}|},
\label{eq:IWDR}
\end{equation}
where $\Omega$ is the rotation rate and $k_\parallel$ the
component of the wave vector ${\bf k}$ along the rotation axis
(referred to as the vertical axis by
convention)~\cite{GreenspanBook}. Accordingly, only fluid motions
at frequencies $\sigma$ smaller than the Coriolis frequency
$2\Omega$ correspond to wave propagation. Fluid motions of weak
amplitude and slowly varying in time ($\sigma \ll 2\Omega$) can be
described in terms of waves with nearly horizontal wave vectors:
they tend to be two-dimensional and three-component (2D3C),
invariant along the rotation axis, a result known as the
Taylor-Proudman theorem.

The trend towards two-dimensionality is a landmark in rotating
turbulence, observed both in experiments and numerical
simulations~\cite{Hopfinger1982,Jacquin1990,Bartello1994,Moisy2011,DavidsonBook2013,Godeferd2015}.
It originates from the modification of the nonlinear interactions
by the Coriolis force, which yields preferential energy transfers
towards modes with almost horizontal wave
vectors~\cite{Morinishi2001,Mininni2009,Lamriben2011,Delache2014}.
In the frequency domain, this corresponds to the generation of
slow dynamics compared to the characteristic frequency at which
energy is supplied to the system. These anisotropic energy
transfers can be accounted for in terms of resonant and
near-resonant triadic interactions of inertial
waves~\cite{Cambon1989,Waleffe1993,Smith1999,Bordes2012}. A major
feature of rotating turbulence is the emergence of inverse energy
transfers in the horizontal
plane~\cite{Smith1999,Baroud2002,Morize2005,Chen2005,Mininni2009,Sen2012,Yarom2013,Deusebio2014,Campagne2014}.
Inverse transfers between 3D fast ``wave'' modes and the 2D slow
``vortex'' mode, mediated by near-resonant triadic interactions,
are allowed at finite Rossby number
only~\cite{Chen2005,Smith2005,Bourouiba2007,Bourouiba2012,Sen2012}.
The 2D mode is therefore fed either from the coupling with the 3D
modes at finite Rossby number, or from direct energy input by the
forcing. One naturally expects the energy within this 2D mode to
undergo an inverse energy cascade, similar to that of classical
(non-rotating) 2D turbulence~\cite{Kraichnan1967,Tabeling2002}.

Such coexistence between 2D and 3D flows is relevant to most
experiments and numerical simulations, and cannot be accounted for
by wave turbulence theory, which describes the {\it direct} energy
cascade arising from resonant triadic interactions of 3D wave
modes only~\cite{Galtier2003,Cambon2004}. This theory therefore
provides only a partial description of rotating turbulence in
realistic systems, and careful experimental and numerical studies
remain necessary to assess its range of validity.

\begin{figure}
\centerline{\includegraphics[width=8cm]{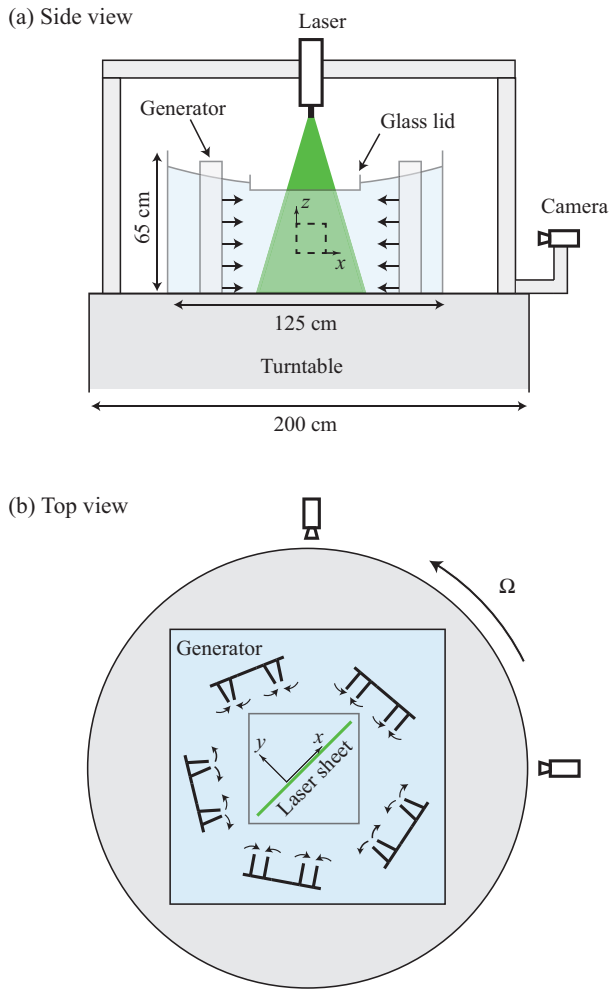}}
\caption{(Color online) Experimental setup. An arena of 10 pairs
of flaps forces a turbulent flow in the central region of a water
tank mounted on a rotating turntable. A laser sheet illuminates a
vertical slice through a horizontal glass lid covering the fluid.
2D3C (two-dimensional three-component) velocity measurements are
performed using stereoscopic PIV (particle image velocimetry) in a
vertical square domain of size $\Delta x \times \Delta z = 14
\times 14$~cm$^2$, shown as a dashed square in panel
(a).}\label{fig:su}
\end{figure}

\begin{figure}
\centerline{\includegraphics[width=7cm]{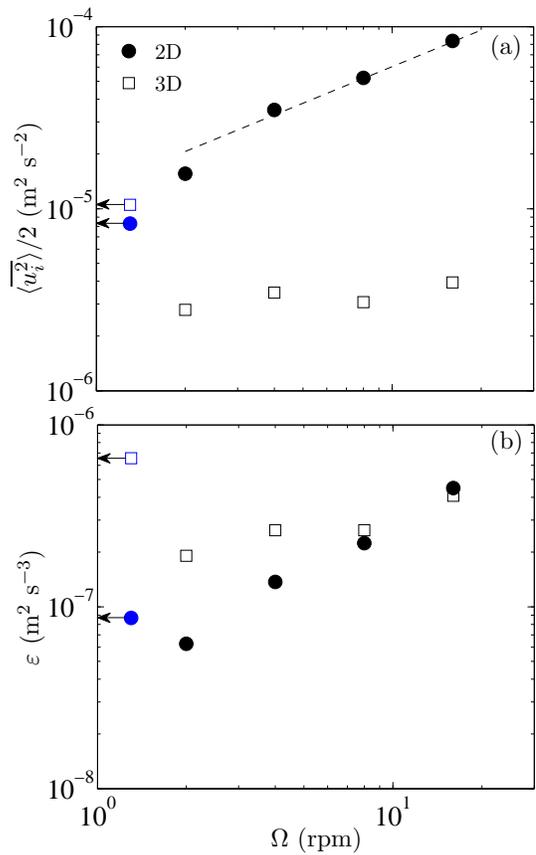}}
\caption{(Color online) (a) Energy and (b) energy dissipation rate
per unit mass for the 2D and 3D modes as a function of the
rotation rate $\Omega$. In both figures, the first data points
(shown with arrows, at arbitrary abscissae) correspond to the
non-rotating case $\Omega=0$.}\label{fig:evsw}
\end{figure}

{Laboratory experiments differ from most numerical and
theoretical studies by the presence of rigid horizontal
boundaries, where the rotating flow achieves no-slip conditions
through Ekman layers \cite{Pedlosky}. In a laminar Ekman layer,
the balance between the viscous and Coriolis forces leads to a
boundary layer thickness $\delta_{Ek}\simeq \sqrt{\nu/\Omega}$.
The belief is that, provided the experimental tank is deep enough,
the  bulk turbulent flow away from the top and bottom boundaries
should resemble the one obtained in the ideal 3D periodic or
stress-free domains considered in most numerical and theoretical
studies. Closer to the horizontal walls, the boundary layers
induce Ekman friction that is not taken into account by most
numerical studies.}

{In the laboratory, the energy dissipation of rotating
turbulent flows originates from three main contributions: bulk
viscous dissipation of 3D flow structures, bulk viscous
dissipation of quasi-2D flow structures (somewhat similar to the
bulk energy dissipation of 2D turbulence), and dissipation through
Ekman friction on the horizontal boundaries.}

In spite of the importance of the 2D mode in most geophysical and
laboratory flows, the 3D fluctuations still play a crucial role in
the dynamics of rotating turbulence, because they are much more
efficient at dissipating energy. This key feature is illustrated
in Fig.~\ref{fig:evsw}, from data obtained in the present
experiment (setup sketched in Fig.~\ref{fig:su}; see
Sec.~\ref{secEbudget} for details): we decompose the turbulent
velocity field into a vertically-averaged 2D flow and a
vertically-dependent 3D remainder, and show the corresponding
energies and energy dissipation rates as a function of global
rotation. For maximum rotation, although the 3D component contains
a small fraction of the total kinetic energy, its dissipation rate
is as large as that of the vertically-invariant 2D component.
{Moreover, both of these dissipations are larger than an
estimate of the frictional losses due to laminar Ekman layers (see
Sec.~\ref{secEbudget})}. An accurate description of the 3D
structures of the flow is therefore essential to characterize the
energy fluxes in rotating turbulence at moderate Rossby number.

A primary goal in this direction is to determine the range of
scales and frequencies for which 3D fluctuations follow the
inertial-wave dispersion relation. This requires a full
spatio-temporal analysis, which is very demanding in general for
wave-turbulence systems: the accessible range of scales is usually
limited in experiments, whereas long integration times are
prohibitive in numerical simulations. The case of rotating
turbulence is particularly delicate because of the specific form
of the dispersion relation (\ref{eq:IWDR}): the frequency is not
related to the wave number, as in conventional isotropic wave
systems such as surface waves~\cite{Berhanu2013} or elastic
waves~\cite{Boudaoud2008,Mordant2008}, but to the wave vector
orientation only.

The recent studies of Clark di Leoni {\it et al.}~\cite{Clark2014}
and Yarom and Sharon~\cite{Yarom2014} constitute important steps
forward in this respect. Using numerical simulation of
rotating turbulence forced at large scale, Clark di Leoni {\it et
al.}~\cite{Clark2014} observe a clear concentration of energy
along the dispersion relation of inertial waves, and provide a
detailed analysis of the various time scales of the system. They
observe a wave-dominated regime at large scale and a
sweeping-dominated regime at small scale (see also
Ref.~\cite{Favier2010}). In the experiment of Yarom and
Sharon~\cite{Yarom2014} the forcing consists of a random set of
sources and sinks at the bottom of a rotating water tank. They
measure three-dimensional two-component (3D2C) velocity fields
using a scanning particle image velocimetry (PIV) technique, and
observe also a good agreement with the  inertial wave dispersion
relation. In both Refs.~\cite{Clark2014} and \cite{Yarom2014}, the
inertial waves are observed at scales smaller than the injection
scale, suggesting that they are fed by forward energy
transfers, which is consistent with the predictions of
wave turbulence theory.

The aim of the present paper is to further analyze experimentally
the range of spatio-temporal scales at which inertial waves can be
detected in rotating turbulence. Stationary rotating turbulence is
produced by a set of vortex dipole generators which continuously
inject turbulent fluctuations towards the center of a rotating
water tank where measurements are performed. We showed in
Ref.~\cite{Campagne2014} that this configuration generates a
double energy cascade at large rotation rate: an inverse cascade
of horizontal energy and a direct cascade of vertical energy,
which behaves approximately as a passive scalar advected by the
horizontal flow. Here we perform a detailed spatio-temporal
analysis using two-point spatial correlations of the temporal
Fourier modes computed from time-resolved two-dimensional
three-component (2D3C) velocity fields measured by stereoscopic
PIV in a vertical plane. We observe that, at large scales and
frequencies, the spatio-temporal anisotropy of the energy
distribution is well described by the dispersion relation of
inertial waves, whereas smaller scales are dominated by the
sweeping of the waves by the energetic large-scale flow.

\section{Experimental setup}

The experimental setup, sketched in Fig.~\ref{fig:su}, is similar
to the one described in Refs.~\cite{Campagne2014,Gallet2014} and
is only briefly described here. It consists of a
$125\times125\times65$~cm$^3$ glass tank, filled with $H=50$~cm of
water and mounted on a 2~m-diameter rotating platform which
rotates at a rate $\Omega$ in the range $0.21$ to
$1.68$~rad~s$^{-1}$ ($2$ to $16$~rpm). Turbulence is produced in
the rotating frame by a set of ten vertical vortex dipole
generators organized as a circular arena of 85~cm diameter around
the center of the water tank.  This forcing device was initially
designed to generate turbulence in stratified fluids, and is
described in detail in Refs.~\cite{Billant2000,Augier2014}. Each
generator consists of a pair of vertical flaps, 60~cm high and
$L_f=10$~cm long, alternatively closing rapidly and opening slowly
in a cyclic motion of period $T_{0}=2\pi/\sigma_{0}=8.5$~s. The
closing stage is achieved with the flaps rotating at an angular
velocity $\sigma_f=0.092$~rad~s$^{-1}$, and a random phase shift
is applied between the generators.

In the laminar regime, a single pair of flaps generates vortex
dipoles with core vorticity $\omega_f$. Additional PIV
measurements close to a vortex dipole generator indicate that this
core vorticity is governed by the vorticity in the viscous
boundary layer of the flap, $\omega_f \sim \sigma_f L_f/\delta$,
where $\sigma_f L_f$ is the flap velocity and $\delta$ the viscous
boundary layer thickness. In the present experiment, the vortex
dipoles are unstable, and the closing of the flaps therefore
produces small-scale 3D turbulent fluctuations that are advected
towards the center of the arena by the remaining large-scale
dipolar structure.

The turbulent Reynolds number, computed from the rms velocity and
the horizontal integral scale, is about 400 in the center of the
flow, and the turbulent Rossby number covers the range $0.30-0.07$
for $\Omega = 2-16$~rpm~\cite{Campagne2014}.

We measure the three components of the velocity field ${\bf
u}=u_x{\bf e}_x+ u_y{\bf e}_y+ u_z{\bf e}_z$ (with ${\bf e}_z$
oriented vertically, along the rotation axis) in a vertical square
domain of size $\Delta x \times \Delta z = 14 \times 14$~cm$^2$
located at the center of the circular arena at mid-depth, using a
stereoscopic PIV system~\cite{DaVis,Pivmat} embarked on the
rotating platform. These 2D3C (two-dimensional three-component)
velocity fields are sampled on a grid of $80\times 80$~vectors
with a spatial resolution of $1.75$~mm. Two acquisition sets are
recorded for each rotation rate $\Omega$: one set of 10\,000
fields at 0.35~Hz and one set of 1\,000 fields at 1.5~Hz. The
combination of these two time series results in a temporal
spectral range of three decades.

\section{2D vs. 3D flow components}\label{secEbudget}

In the present experiment, energy is primarily injected in the 2D
mode (vertically invariant), but the instabilities in the vicinity
of the flaps rapidly feed 3D fluctuations which are advected in
the central region. Energy transfers between the 2D and 3D flow
components, which vanish in the weak turbulence limit ($Ro
\rightarrow 0$), are allowed in our system because of the moderate
value of the Rossby number. It is therefore of first interest to
quantify the energy contained in the 2D and 3D components of the
flow. We estimate the vertically-averaged 2D flow as the average
of the velocity field over the vertical extent $\Delta z$ of the
PIV field,
\begin{equation}
{\bf u}_{2D} = \frac{1}{\Delta z} \int_0^{\Delta z} {\bf u}
(x,z) dz, \label{eq:u2d}
\end{equation}
and the remaining $z$-dependent 3D flow as ${\bf u}_{3D} = {\bf u}
- {\bf u}_{2D}$. We compute the energy per unit mass of these two
flow components as $\langle \overline{{\bf u}_{2D}^2} \rangle/2$
and $\langle \overline{{\bf u}_{3D}^2} \rangle/2$, with
$\overline{\,\cdot\,}$ the temporal average and $\langle \,\cdot\,
\rangle$ the spatial average over the PIV field. They are plotted
in Fig.~\ref{fig:evsw}(a) as a function of the rotation rate
$\Omega$. Because of the limited height of the PIV field, the 2D
flow estimated from Eq.~(\ref{eq:u2d}) unavoidably contains 3D
fluctuations associated to vertical scales larger than $\Delta z$,
so the measured 2D energy may overestimate the true one.

Figure~\ref{fig:evsw}(a) shows that without rotation the 2D and 3D
components of the flow have comparable energy. With rotation, the
2D energy increases with $\Omega$, following approximately the
power law $\Omega^{2/3}$~\cite{Gallet2014}, whereas the 3D energy
remains approximately constant, and represents only 5\% of the
total energy at the largest rotation rate. Although most of the
energy is contained in the 2D flow component for $\Omega \neq 0$,
a significant fraction of the dissipation still arises from the 3D
fluctuations. Assuming axisymmetry, we compute an estimate of the
energy dissipation rate $\epsilon = \nu \langle
\overline{(\partial u_i / \partial x_j)^2} \rangle$ from the 6
terms of the velocity gradient tensor accessible in the 3C2D
measurements,
\begin{equation}
\begin{multlined}
\epsilon \simeq \nu \left<
  2 \overline{\left(\frac{\partial u_x}{\partial x}\right)^2}
+ 2 \overline{\left(\frac{\partial u_y}{\partial x}\right)^2}
+ 2 \overline{\left(\frac{\partial u_z}{\partial x}\right)^2} \right.\\
\left. + \overline{\left(\frac{\partial u_x}{\partial z}\right)^2}
+ \overline{\left(\frac{\partial u_y}{\partial z}\right)^2}
+ \overline{\left(\frac{\partial u_z}{\partial z}\right)^2}
\right> .
\end{multlined}
\end{equation}
This dissipation rate, computed both for ${\bf u}_{2D}$ and ${\bf
u}_{3D}$, is shown in Fig.~\ref{fig:evsw}(b). Since the
derivatives are obtained from finite differences at the smallest
resolved scale, the computed dissipation underestimates the true
one (the PIV resolution is $1.75$~mm while the Kolmogorov scale is
of order of 0.6~mm~\cite{Gallet2014}). However, we expect the
measured evolution of $\epsilon$ with $\Omega$ to reflect the true
one.

{We first compare these bulk energy dissipation rates to an
estimate of frictional losses due to laminar Ekman layers,
$\epsilon_{Ek}\simeq \nu \frac{U_{\perp
\text{rms}}^2}{\delta_{Ek}^2} \frac{\delta_{Ek}}{H}= \sqrt{\nu
\Omega} \frac{U_{\perp \text{rms}}^2}{H}$, where $U_{\perp
\text{rms}}$ is the root-mean-square horizontal velocity. This
estimate ranges from $8\times10^{-9}$~m$^2$~s$^{-3}$ for
$\Omega=2$ rpm to $1\times10^{-7}$~m$^2$~s$^{-3}$ for $\Omega=16$
rpm: it is smaller than the bulk energy dissipation of both the 2D
and 3D parts of the turbulent flow, by a factor of 10 for slow
rotation and 4 for rapid rotation. A detailed experimental
characterization of these Ekman layers would be necessary to
validate the assumption of laminar layers, but it is beyond the
scope of the present study.}

{We now compare the bulk energy dissipation rates in the 2D
and 3D parts of the turbulent flow.} Remarkably, while the 3D
fluctuations represent a small fraction of the total energy, they
account for a large fraction of the dissipation at all rotation
rates. It is therefore of first interest to investigate these 3D
modes, and to determine to what extent they can be described in
terms of inertial waves.

\section{Temporal analysis}

We now focus on the temporal dynamics of the velocity field, which
we characterize through the energy distribution of turbulent
fluctuations as a function of angular frequency $\sigma$. This
temporal energy spectrum is defined as
\begin{equation}\label{eq:Es}
E(\sigma) = \frac{4\pi}{T}\left\langle|\tilde{u}_{i}({\bf
x},\sigma)|^2\right\rangle,
\end{equation}
where
\begin{equation}
\tilde{u}_{i}({\bf x},\sigma)=\frac{1}{2\pi}\int_{0}^{T} u_i({\bf
x},t)e^{-i\sigma t}\,dt
\end{equation}
is the temporal Fourier transform of the velocity field
${u}_{i}({\bf x},t)$ (with $i=x,y,z$), $T$ the acquisition
duration and $\langle\,\cdot\,\rangle$ the spatial average. The
normalization is such that $\langle\overline{ u_i^2}\rangle =
\int_0^\infty E(\sigma)\,d\sigma$, with $\overline{\,\cdot\,}$ the
temporal average. We use the standard Welch's
method~\cite{Welch1967} to improve the statistical convergence of
the power spectrum.

\begin{figure}
\centerline{\includegraphics[width=8.5cm]{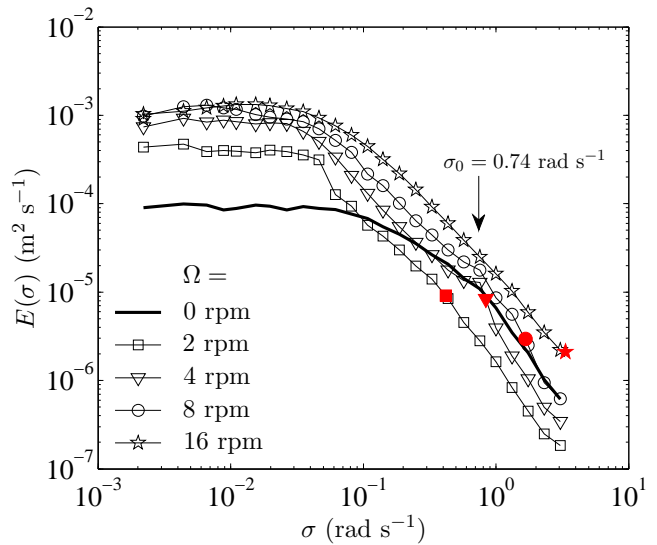}}
\caption{(Color online) Temporal energy spectrum $E(\sigma)$ as a
function of the angular frequency $\sigma$ for different rotation
rates $\Omega$. $\sigma_{0}$ indicates the frequency of the
opening-and-closing cycle of the flaps. The Coriolis frequency
$2\Omega$ is highlighted with filled symbols.}\label{fig:esigma}
\end{figure}

We plot $E(\sigma)$ for each rotation rate $\Omega$ in
Fig.~\ref{fig:esigma}. For $\Omega\neq 0$, we observe a global
increase with $\Omega$ of the energy at all frequencies,
consistently with the behavior of the overall energy in
Fig.~\ref{fig:evsw}(a). These spectra for $\Omega\neq 0$ strongly
differ from the non-rotating spectrum, with relatively much more
energy at low frequency in the rotating case: global rotation
induces slow dynamics.

\begin{figure}
\centerline{\includegraphics[width=8.3cm]{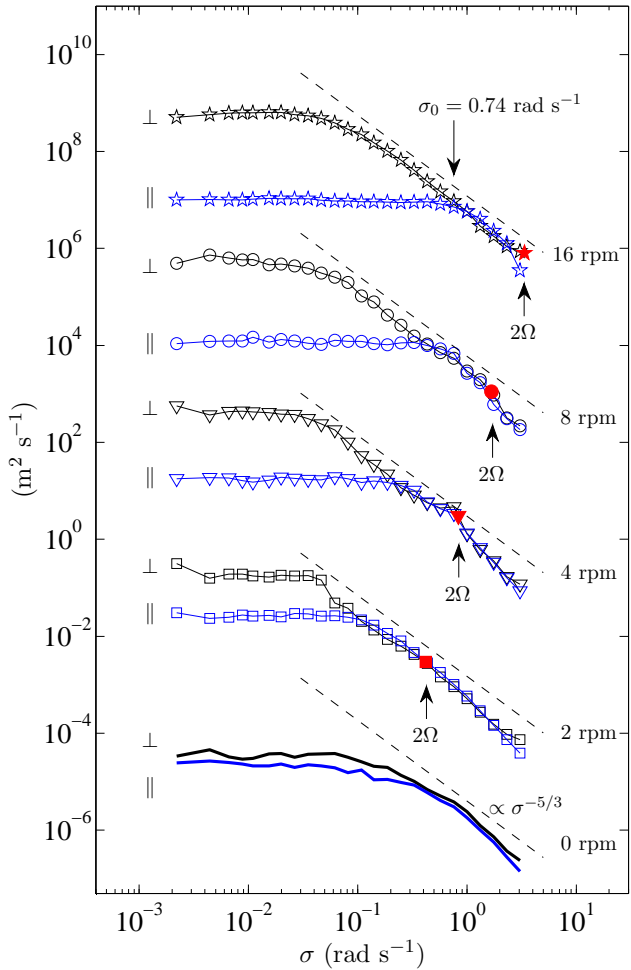}}
\caption{(Color online) Temporal energy spectra of the vertical,
$E_\parallel$ (light gray, blue), and horizontal, $E_\perp/2$
(black), velocity components for different rotation rates $\Omega$.
For visibility, there is a vertical shift by a factor of $10^3$
between couples of curves at different $\Omega$. The dashed lines
show power laws $\sigma^{-5/3}$. }\label{fig:eperppar}
\end{figure}

A first step towards a description of the flow anisotropy in the
frequency domain can be provided by further decomposing the power
spectrum density~(\ref{eq:Es}) as
\begin{equation}\label{eq:EEE}
E(\sigma) = E_\parallel(\sigma) + E_\perp(\sigma),
\end{equation}
with $E_\parallel(\sigma) = 4\pi \langle |\tilde u_z({\bf
x},\sigma)|^2 \rangle/T$ the spectrum of the vertical velocity and
$E_\perp(\sigma) = 4\pi\langle |\tilde u_x({\bf x},\sigma)|^2
\rangle/T + 4\pi\langle |\tilde u_y({\bf x},\sigma)|^2 \rangle/T$
the spectrum of the horizontal velocity. This decomposition
highlights the frequency-dependent {\it componentality} of
turbulence, i.e., the distribution of energy among the different
velocity components, which is related to the {\it polarization
anisotropy}~\cite{Cambon1989,Morinishi2001,Delache2014}. This is
not to be confused with the frequency-dependent {\it
dimensionality} of turbulence, which compares the vertical and
horizontal characteristic scales at a given frequency (described
in section~\ref{sec:sta}).

The temporal spectra $E_\parallel(\sigma)$ and $E_\perp(\sigma)/2$
are shown in Fig.~\ref{fig:eperppar} for all rotation rates.
Without rotation, energy is nearly equally distributed among the 3
velocity components (i.e., $E_\perp \simeq 2 E_\parallel$). There
 is actually a slight over-representation of horizontal energy, a
consequence of the forcing device geometry which preferentially
injects energy in horizontal motions. As the rotation rate
increases, the high frequencies remain nearly isotropic
(iso-component), whereas the low frequencies become gradually
anisotropic, with $E_\parallel(\sigma)$ nearly flat and
$E_\perp(\sigma)$ approaching a power law close to
$\sigma^{-5/3}$. This anisotropy is related to the fact that, as
$\Omega$ increases, the decorrelation frequency (i.e., the
frequency below which the spectrum becomes flat) gets
significantly smaller for the horizontal velocity
($\sigma_{\text{dec}\perp}=0.04 \pm 0.01$) than for the vertical
velocity ($\sigma_{\text{dec}\parallel}$, increasing from $0.05$
to $0.40$~rad~s$^{-1}$ for $\Omega$ from $2$ to $16$~rpm).
{For the largest rotation rate ($\Omega=16$~rpm), there is a
clear range of frequencies over which the horizontal spectrum
$E_\perp(\sigma)$ follows a $\sigma^{-5/3}$-power-law. This range
gets narrower for decreasing rotation rate $\Omega$.}

\begin{figure}
\centerline{\includegraphics[width=8.3cm]{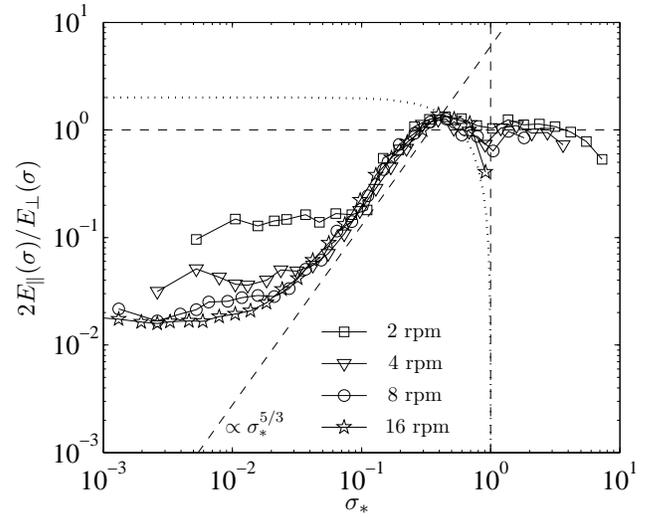}}
\caption{Componential anisotropy ratio as a function of the
normalized frequency $\sigma_{*} = \sigma / 2\Omega$. Isotropy is
indicated by the horizontal dashed line at $2 E_\parallel /
E_\perp = 1$. The dotted line indicates the prediction for a plane
inertial wave, i.e., $2E_\parallel /
E_\perp=2(1-\sigma_*^2)/(1+\sigma_*^2)$.}\label{fig:componen}
\end{figure}

The decorrelation frequency of the vertical velocity appears to
scale as $\sigma_{\text{dec}\parallel}* =
\sigma_{\text{dec}\parallel} / 2\Omega = 0.25\pm 0.05$, which
becomes evident when plotting the ratio $2E_\parallel / E_\perp$
as a function of the normalized frequency $\sigma_{*} = \sigma /
2\Omega$ (Fig.~\ref{fig:componen}). In this figure, for
$\sigma_{*}>1$, for which no inertial waves can exist, energy is
nearly equally distributed among the velocity components
($2E_\parallel / E_\perp \simeq 1$). The frequency range over
which a power law $\sigma_*^{5/3}$ is approached is bounded by the
two decorrelation frequencies: on the left by
$\sigma_{\text{dec}\perp}$ and on the right by
$\sigma_{\text{dec}\parallel}$. Interestingly, we also observe a
small frequency domain
$\sigma_{\text{dec}\parallel}*<\sigma_{*}\lesssim0.6$ over which
energy in the vertical component is slightly larger than in each
horizontal component. Such a slight over-representation of the
vertical velocity is compatible with the ``componential''
anisotropy of an assembly of linear inertial waves: a single plane
inertial wave has a componential anisotropy $2E_\parallel /
E_\perp=2(1-\sigma_*^2)/(1+\sigma_*^2)$ (shown as a dotted line in
Fig.~\ref{fig:componen}), with a larger rms velocity along the
vertical than along any horizontal direction for
$\sigma^*<1/\sqrt{3}\simeq 0.6$.

As can be seen in Fig.~\ref{fig:evsw}, for the experiment under
rapid rotation, nearly 90\% of the energy is contained in the 2D
vortex mode. As discussed quantitatively in Campagne~\textit{et
al.}~\cite{Campagne2014}, this strong 2D nature of the flow drives
an inverse cascade of energy for the horizontal velocity and a
direct cascade of energy for the vertical velocity. The horizontal
velocity consequently exhibits slow dynamics, while the vertical
velocity fluctuations are found at higher frequencies
(Fig.~\ref{fig:eperppar}). This behavior is consistent with the
usual phenomenology of rapidly rotating turbulence: {the flow
becomes approximately 2D at low-frequency, and the vertical
velocity behaves as a passive scalar, which is stretched and
folded by the horizontal velocity. This produces thin vertical
layers swept by the horizontal flow, yielding rapidly changing
time series of the vertical velocity. In a similar fashion, the
$\sigma^{-5/3}$-power-law of $E_\perp(\sigma)$ could originate
from the stochastic sweeping by the large scale horizontal flow of
a $k_\perp^{-5/3}$ spatial spectrum, reminiscent of the inverse
energy cascade of 2D turbulence.}

\section{Spatio-temporal analysis}
\label{sec:sta}

\subsection{Spatio-temporal correlations}

We now turn to a combined spatio-temporal analysis of the PIV time
series, focusing on the signature of inertial waves in terms of
{\it dimensional} anisotropy. This signature is sought here in
terms of characteristic horizontal and vertical scales of the
turbulent structures as a function of their frequency. For this,
we define the frequency-dependent two-point spatial correlation of
the temporal Fourier transform of the velocity field
\begin{equation}
R({\bf r},\sigma)= \frac{2\pi}{T} \left\langle\tilde{u}_{i}({\bf
x},\sigma)\tilde{u}_{i}^*({\bf x+r},\sigma) + {\rm
c.c.}\right\rangle, \label{eq:r}
\end{equation}
with $^*$ the complex conjugate (here again Welch's method is used
to improve convergence). Instead of the spectra considered in
Refs.~\cite{Clark2014,Yarom2014}, we compute spatial correlations,
because the former are sensitive to finite size effects arising
from the PIV field being of limited extent compared to the largest
flow structures. The correlation (\ref{eq:r}) probes the energy
distribution among vector separations ${\bf r}$ for each frequency
$\sigma$. Its angular average provides an estimate for the
cumulative energy from scale $r=|{\bf r}|$ to $r=\infty$ for
turbulent motions of frequency $\sigma$. The single-point limit of
this correlation is the temporal energy spectrum (\ref{eq:Es}),
i.e. $E(\sigma)=R({\bf r}=0,\sigma)$.

\begin{figure}
\centerline{\includegraphics[width=9cm]{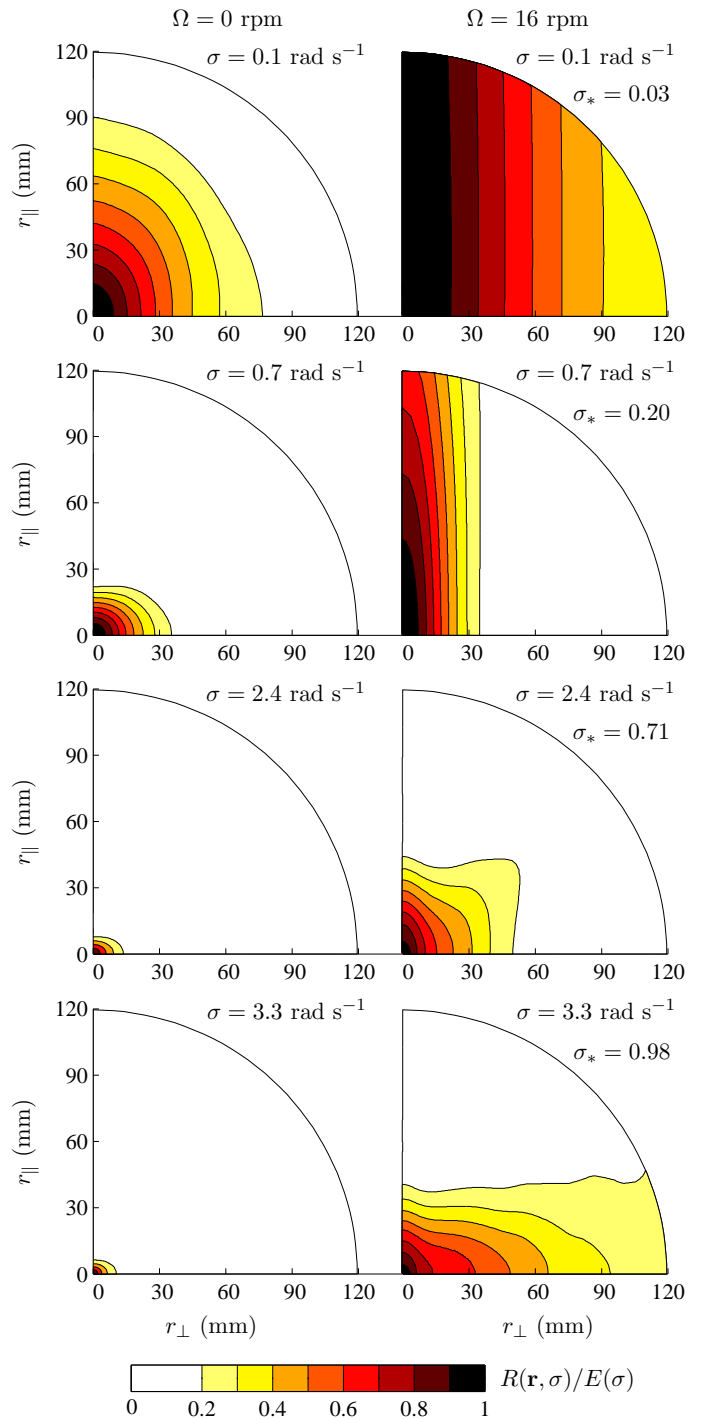}}
\caption{(Color online) Maps of the normalized two-point
correlation $R({\bf r},\sigma)/E(\sigma)$ in the vertical plane
$(r_\perp,r_\parallel)$ for $\Omega=0$ (left) and $\Omega=16$~rpm
(right), at four frequencies $\sigma= 0.1, 0.7, 2.4, 3.3
$~rad~s$^{-1}$. In the rotating case, the corresponding normalized
frequencies are $\sigma_{*}=\sigma/2\Omega = 0.03, 0.20, 0.71,
0.98$. Isocontour lines for $R({\bf r},\sigma)/E(\sigma) < 0.2$
are noisy and are not shown.} \label{fig:mapR}
\end{figure}

Maps of the normalized correlation $R({\bf r},\sigma)/E(\sigma)$
are plotted in Fig.~\ref{fig:mapR} for $\Omega=0$ and 16~rpm at
four selected angular frequencies $\sigma$. In the non-rotating
case, the iso-$R$ lines are approximately circular at all scales
and frequencies, indicating the overall isotropy of turbulence.
The strongly peaked correlation that develops around ${\bf r}=0$
as $\sigma$ increases indicates that rapid turbulent fluctuations
are found at small scales only. In the rotating case, the iso-$R$
lines evolve gradually from quasi-vertical at small frequency
(``cigar'' anisotropy) to more horizontal for $\sigma_{*}\sim 1$
(``pancake'' anisotropy), with $\sigma_{*} = \sigma / 2\Omega$ the
normalized frequency. The ``cigar'' anisotropy observed at
$\sigma_{*}\ll 1$ is consistent with the 2D3C vertical invariance
predicted by the Taylor-Proudman theorem for vanishing frequency:
it corresponds to the zero-frequency limit of Eq.~(\ref{eq:IWDR})
for nearly horizontal wave vector. Similarly, the tendency towards
``pancake'' anisotropy, observed for $\sigma_{*} \rightarrow 1$
and sufficiently large $r_\perp$, is consistent with the nearly
vertical wave vector limit of Eq.~(\ref{eq:IWDR}).

A natural way to characterize the frequency-dependent anisotropy
would be to compute integral scales along and normal to the
rotation axis at each frequency. Here we consider a finer
approach, which also takes into account the scale-dependence of
this anisotropy: for each frequency $\sigma$ and horizontal scale
$r_\perp$, we identify the vertical scale
$\ell_\parallel(r_\perp,\sigma)$ at which the correlation along
the vertical axis is equal to the one at $r_\perp {\bf e}_\perp$,
i.e. such that $R({\bf r} = \ell_\parallel {\bf e}_\parallel,
\sigma)=R({\bf r} = r_\perp {\bf e}_\perp, \sigma)$. In practice,
we compute $\ell_\parallel$ as the vertical semi-axis obtained
from the fit of the iso-$R$ line defined by $R({\bf
r},\sigma)=R(r_\perp {\bf e}_\perp,\sigma)$ with an ellipse of
prescribed horizontal semi-axis $r_\perp$. This method allows us
to filter out the noise in the iso-$R$ lines at small $R$. It also
extends the analysis to values of $\ell_\parallel$ larger than the
PIV field height ($\Delta z = 140$~mm), which is useful at small
$\sigma_{*}$ for nearly vertically invariant $R$. Poor fits
defined by a correlation coefficient less than 0.9 or such that
$\ell_\parallel$ is larger than $2 \Delta z$ are discarded. We
finally define the scale and frequency dependent anisotropy factor
as
\begin{equation}
A(r_\perp,\sigma) = \frac{r_\perp}{\ell_\parallel(r_\perp,\sigma)}.
\label{eq:a}
\end{equation}
It is equal to $1$ for isotropic turbulence, to $0$ for vertically
invariant (2D3C) turbulence, and to $\infty$ for horizontally
invariant (1D2C) turbulence.

If the anisotropy of the two-point correlation $R$ at frequency
$\sigma_{*}\leqslant 1$ is governed by linear inertial waves, we
expect $A$ to be independent of the scale, and to be set by the
dispersion relation~(\ref{eq:IWDR}). A simple estimate, assuming
that the wave vector ${\bf k}$ is in the vertical measurement
plane and identifying  on dimensional grounds $\ell_\parallel \sim
k_\parallel^{-1}$ and $r_\perp \sim k_\perp^{-1}$, yields a
frequency-dependent anisotropy factor $A_{IW}(\sigma_{*}) \simeq
(\sigma_*^{-2}-1)^{-1/2}$. Considering now an assembly of inertial
waves with an axisymmetric wavevector distribution, an analytic
computation (given in the Appendix) leads to a very similar
result,
\begin{equation}\label{eq:aiw}
A_{IW}(\sigma_{*})=\sqrt{\frac{2}{\sigma_*^{-2}-1}}.
\end{equation}

\begin{figure}
\includegraphics[width=8.3cm]{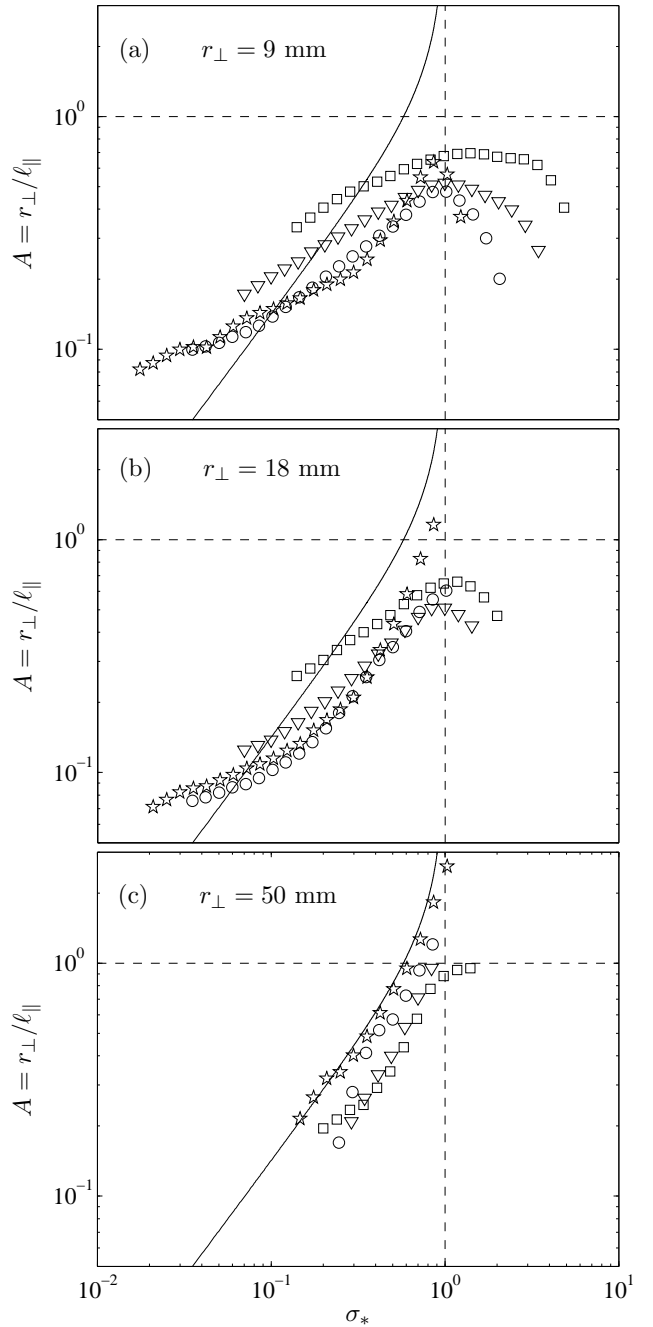}
\caption{Anisotropy factor $A$ (\ref{eq:a}) as a function of the
normalized frequency $\sigma_{*}=\sigma/2\Omega$ at different
rotation rates (same symbols as in Fig.~\ref{fig:esigma}), for
three horizontal scales $r_\perp = 9$,~$18$ and $50$~mm. The
continuous line represents the inviscid inertial-wave prediction
$A_{IW}$ (\ref{eq:aiw}).}\label{fig:aniso}
\end{figure}

In Figs.~\ref{fig:aniso}\,(a-c), we compare the anisotropy factor
$A$ measured in the four rotating experiments to the inertial-wave
prediction (\ref{eq:aiw}) at three horizontal scales $r_\perp =
9$,~$18$ and $50$~mm. For all scales and $\sigma_{*}\leq 1$, $A$
is an increasing function of $\sigma_{*}$, confirming that slow
fluctuations are more vertically elongated than fast fluctuations.
We find that the inertial-wave prediction (\ref{eq:aiw}) provides
a good description of the data at large horizontal scales and
large rotation rate. For such large scales ($r_\perp \simeq
50$~mm), the anisotropy factor is no longer accessible for
$\sigma_{*}<0.1$ because it corresponds to $\ell_\parallel$ much
larger than the height of the PIV field. On the other hand, at
smaller horizontal scale the prediction (\ref{eq:aiw}) fails, with
small frequencies more isotropic than predicted by the
inertial-wave argument.

Because of the moderate Reynolds and Rossby numbers of the present
experiment, two effects may be considered to explain why large
scales follow the inertial-wave prediction whereas small scales do
not: viscous damping and sweeping of small scales by the velocity
at larger scales. Viscosity introduces an imaginary term $i \nu
|{\bf k}|^2$ in the dispersion relation (\ref{eq:IWDR}) without
modifying its real part. Waves $(\sigma,{\bf k})$ such that $|{\bf
k}| \gg r_\nu^{-1}$ are therefore damped, with $r_\nu =
\sqrt{\nu/\sigma}$ a viscous cutoff. This viscous cutoff is of
order of 1 to 10~mm for the normalized frequencies $\sigma_*=
\sigma/2\Omega$ in the range $[10^{-2},1]$ considered in
Fig.~\ref{fig:componen}. However, since viscous damping affects
the wave amplitude without modifying the wave vector components,
its should not affect the anisotropy. We therefore focus in the
following on the sweeping effect.

\subsection{Sweeping effect}

Sweeping corresponds to the advection of the waves by the
large-scale flow, which leads to a modification of their apparent
frequency. An inertial wave propagating in a time-independent
uniform flow ${\bf U}$ has a Doppler-shifted frequency,
\begin{equation}
 \sigma=\sigma_i + {\bf k} \cdot {\bf U} \, ,\label{eq:sweptIWDR}
 \end{equation}
where $\sigma_i$ is the intrinsic frequency given by
(\ref{eq:IWDR}), and $\sigma$ is the frequency at which the wave
is detected in the frame of the rotating tank. In our experiment,
the energetic large-scale 2D flow may be thought of locally as a
uniform sweeping flow $\bf U$ that evolves slowly in time,
inducing a scrambling of the waves' spatio-temporal signature. An
order of magnitude of the typical Doppler-shift can be estimated
by $k_\perp U_{\perp \text{rms}}$, where $U_{\perp \text{rms}}$ is
the root-mean-square horizontal velocity. A key difference between
equations (\ref{eq:sweptIWDR}) and (\ref{eq:IWDR}) is that the
frequency $\sigma$ now depends on the magnitude of ${\bf k}$, with
small-scale waves more affected by sweeping.

For an ensemble of inertial waves with axisymmetric wavevector
statistics, the intrinsic frequency $\sigma_i$ can be related to
the anisotropy through Eq.~(\ref{eq:aiw}). Subsituting the
corresponding expression into (\ref{eq:sweptIWDR}) and estimating
the Doppler-shift term on dimensional grounds, we obtain
\begin{equation}\label{eq:sweff}
\sigma \simeq \frac{2 \Omega}{\sqrt{1+2A^{-2}}} + {\cal C}
\frac{U_{\perp \text{rms}}}{r_\perp}\, ,
\end{equation}
where ${\cal C}$ is a constant of order unity.
This indicates that the parameter
 \begin{equation}
N = \frac{2 \Omega r_\perp }{U_{\perp \text{rms}} \sqrt{1+\frac{2}{A^2}}}
\label{eq:scaling}
 \end{equation}
should be a unique function of the sweeping parameter $S =
U_{\perp \text{rms}} / \sigma r_\perp$. $N$ corresponds
approximately to the intrinsic frequency of inertial waves
rescaled by the advective time $r_\perp/U_{\perp \text{rms}}$,
whereas $S$ is the observed period of the waves, rescaled by the
same advective time.

\begin{figure}
\centerline{\includegraphics[width=8cm]{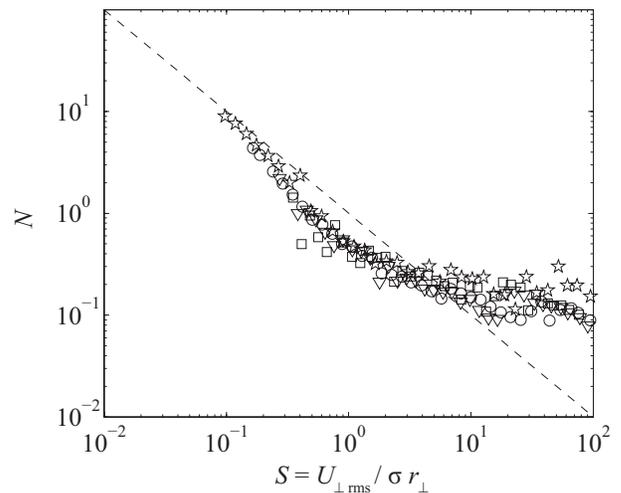}}
\caption{Rescaled intrinsic frequency $N$ (\ref{eq:scaling}) as a
function of the sweeping parameter $S$. The symbols indicate the
different rotation rates, and are the same as in
Fig.~\ref{fig:esigma}. The dashed line $N \sim 1/S$ shows the
low-$S$ prediction for (non-swept) ensembles of inertial waves
[Eq.~(\ref{eq:scaling})]. For $S \ll 1$ one sees the signature of
the dispersion relation (\ref{eq:IWDR}), while for $S \gg 1$ one
detects swept inertial waves.} \label{fig:collaps}
\end{figure}

Figure~\ref{fig:collaps} confirms this picture: the data for
different values of $\Omega$, $r_\perp$ and $\sigma$ collapse onto
a master curve $N=f(S)$. This collapse indicates that sweeping is
indeed responsible for the departure from the inertial wave
prediction at small frequencies and/or small scales. The expected
asymptotic behavior for small sweeping parameter is $N \simeq
1/S$, which corresponds to the prediction (\ref{eq:aiw}) for an
axisymmetric ensemble of non-swept inertial waves. The data is in
quantitative agreement with this small-$S$ prediction, shown as a
dashed line in figure~\ref{fig:collaps}. For large $S$, equation
(\ref{eq:sweff}) indicates that $N$ should asymptote to a constant
value $N \simeq {\cal C}$, which again is compatible with the
data.

The master curve in figure~\ref{fig:collaps} has the following
simple interpretation: high-frequency or large-scale waves are
hardly affected by  sweeping. The Doppler-shift term is negligible
compared to their intrinsic frequency, and their location in a
space-time energy distribution is given by the dispersion relation
(\ref{eq:IWDR}). This is the low-$S$ behavior in
figure~\ref{fig:collaps}.

By contrast, when focusing on low frequencies $\sigma$ or
small-scales in the frame of the tank, one measures the inertial
waves with intrinsic frequency $\sigma_i=\sigma$, but one also
detects many waves with $\sigma_i \neq \sigma$ that are
Doppler-shifted back to frequency $\sigma$ by the advective term
in (\ref{eq:sweptIWDR}). The anisotropy measured at low-frequency
$\sigma$ therefore results from strongly swept inertial waves with
various intrinsic frequencies, and the information from the
dispersion relation (\ref{eq:IWDR}) is lost in the space-time
correlation. The limit $S \gg 1$ corresponds to frequencies
$\sigma$ that are much lower than the inverse advective time. In
this $\sigma \to 0$ limit, one detects mostly waves with $\sigma_i
\gg \sigma$ that are Doppler-shifted by the horizontal flow in
such a way that they are almost steady in the frame of the tank:
this is a $\sigma$-independent regime that corresponds to the
large-$S$ plateau in figure~\ref{fig:collaps}.

\section{Conclusion}

In the present experiment, the anisotropy of the turbulent energy
distribution at a given spatio-temporal scale $(r_\perp,\sigma)$
is well-described by the inertial-wave dispersion relation at
high-frequency and/or large-scale only. The smaller-scale waves
are subject to intense sweeping by large-scale turbulent motions
contained predominantly in the 2D ``vortex" mode. This conclusion
is compatible with the numerical findings of Clark di Leoni {\it
et al.}~\cite{Clark2014}, who also identify the sweeping time
scale as the relevant decorrelation time at small scale.

Such sweeping by the 2D mode has strong implications for
wave-turbulence theories. Indeed, most waves do not follow the
inertial-wave dispersion relation, and the assumptions of weak
turbulence theory break down even at the linear stage in wave
amplitude: instead of the dispersion relation (\ref{eq:IWDR}), the
linear problem consists in  determining the evolution of waves
embedded in a turbulent 2D flow. This is a formidable task in
general, because the 2D flow is space- and time-dependent: in the
discussion of our data, we simplified the problem by assuming that
the 2D flow is at much larger scales and slower frequencies than
the waves, therefore including it as a simple Doppler-shift term
in the dispersion relation.

{We conclude with a discussion on the dimensionality of the
forcing. In the present experiment, the flow is driven by
vertically invariant flaps: such a quasi-2D forcing device
enhances two-dimensionalization and the resulting sweeping of the
3D flow structures. Nevertheless, accumulation of energy in the 2D
mode is a robust feature of rotating turbulence, that takes place
for arbitrary forcing geometry, even if the forcing does not input
energy directly into the 2D mode. A careful and extensive
numerical study of this issue has been recently reported for the
fully-3D Taylor-Green forcing \cite{Alexakis}: for rapid global
rotation and low viscosity, energy accumulates in the 2D mode
until the Rossby number based on the velocity of this 2D flow is
of order unity. If these findings are confirmed, the sweeping of
the most energetic 3D structures would be an inevitable outcome of
this accumulation of 2D energy.}

\acknowledgments

We acknowledge P. Augier, P. Billant and J.-M. Chomaz for kindly
providing the flap apparatus, and A. Aubertin, L. Auffray, C.
Borget and R. Pidoux for their experimental help. This work is
supported by the ANR grant no. 2011-BS04-006-01 ``ONLITUR''. B.G.
acknowledges support from Labex PALM. F.M. acknowledges the
Institut Universitaire de France.

\appendix

\section{Anisotropy factor for a statistically axisymmetric distribution of inertial waves}

We compute the anisotropy factor $A$ for an ensemble of
independent plane inertial waves, with axisymmetric wave vector
statistics. The temporal Fourier transform of the velocity field
reads
\begin{equation}
\tilde{{\bf u}}({\bf x},\sigma)= \int {\bf a}({\bf k},{\sigma}) e^{i {\bf k}\cdot {\bf x}} \mathrm{d} {\bf k} \, ,
\end{equation}
where ${\bf a}({\bf k},\sigma)$ is the space-time Fourier
amplitude of the velocity field at wave number ${\bf k}$ and
frequency $\sigma$. The two-point velocity
correlation at frequency $\sigma$ (\ref{eq:r}) can be written
\begin{eqnarray}
\nonumber R({\bf r},\sigma) & = & \frac{1}{2} \left< \iint {\bf a}({\bf k}_1,{\sigma}) \cdot  {\bf a}^*({\bf k}_2,{\sigma}) \right. \\
\nonumber & \times & \left. e^{i ({\bf k}_1\cdot {\bf x}-{\bf k}_2\cdot ({\bf x}+{\bf r}))} \mathrm{d} {\bf k}_1 \mathrm{d} {\bf k}_2  +  \rm{c.c.}  \right> \, ,\\
& = & \int | {\bf a}({\bf k},{\sigma}) |^2 \cos ({\bf k}\cdot {\bf
r}) \mathrm{d} {\bf k} \, ,
\end{eqnarray}
where $\langle \,\cdot\, \rangle$ is the space average, and ${\bf
r}$ is a separation vector inside the PIV plane. Introducing
spherical coordinates with vertical polar axis, we denote
$\varphi$ the azimuthal angle between ${\bf k}$ and the vertical
PIV plane. The argument of the cosine becomes
\begin{eqnarray}
{\bf k}\cdot {\bf r} & = & k_\parallel r_\parallel + k_\perp
r_\perp \cos \varphi \, .
\end{eqnarray}

Let us first consider an ensemble of inertial waves having the
same wavenumber, $|{\bf k}|=k$. For a given reduced frequency
$\sigma_*=\sigma/2 \Omega$, the dispersion relation determines the
ratio $|k_\parallel|/k_\perp$, and because of statistical
axisymmetry $| {\bf a}({\bf k},\sigma) |^2$ is independent of
$\varphi$. The spatial correlation at frequency $\sigma$ becomes
\begin{eqnarray}
\nonumber & R({\bf r},\sigma)  & =   \frac{G(k,\sigma)}{2 \pi} \int_0^{2 \pi} \cos (k_\parallel r_\parallel + k_\perp r_\perp \cos \varphi) \mathrm{d}\varphi \, , \\
 &  = &    G(k,\sigma)  \cos \left(\sigma_* k r_\parallel \right) J_0 \left(\sqrt{1-\sigma_*^2} \, k r_\perp \right),\label{RBessel}
\end{eqnarray}
where $J_0$ is the Bessel function of the first kind and
$G(k,\sigma)$ is a prefactor proportional to the squared amplitude
of the waves at wavenumber $k$ and frequency $\sigma$.

For a given frequency $\sigma$ and horizontal scale $r_\perp$, the
vertical scale $\ell_\parallel(r_\perp,\sigma)$ is determined from
the isolines $R=$constant in the $(r_\perp,r_\parallel)$ plane. An
isoline of $R$ starting on the horizontal axis at $r_\perp$
intersects the vertical axis at
$r_\parallel=\ell_\parallel(r_\perp,\sigma)$. According to
expression (\ref{RBessel}), such isolines connecting the two axes
exist provided the argument of the Bessel function is smaller than
the first zero of this function, i.e.,
\begin{eqnarray}
\sqrt{1-\sigma_*^2} \, k r_\perp < C_0 \, ,\label{criterion}
\end{eqnarray}
where $J_0(C_0)=0$ ($C_0 \simeq 2.40$). Equating expression
(\ref{RBessel}) computed for $(r_\perp,r_\parallel=0)$ and for
$(r_\perp=0, r_\parallel=\ell_\parallel)$ leads to
\begin{eqnarray}
\ell_\parallel(r_\perp,\sigma_*)= \frac{\arccos\left[ J_0
\left(\sqrt{1-\sigma_*^2} \, k r_\perp \right) \right]}{\sigma_*
k},
\end{eqnarray}
and an anisotropy factor
\begin{eqnarray}
A(r_\perp,\sigma_*)= \frac{k r_\perp \sigma_*}{\arccos\left[ J_0 \left(\sqrt{1-\sigma_*^2} \, k r_\perp \right) \right]} \, .
\end{eqnarray}
This anisotropy factor depends very weakly on $k r_\perp$: it is
minimum for low $k r_\perp$, where Taylor expansion for $k r_\perp
\ll 1$ leads to
\begin{eqnarray}
A(r_\perp,\sigma_*) \simeq \sqrt{2} \frac{\sigma_*}{\sqrt{1-\sigma_*^2}} \, ,\label{lowk}
\end{eqnarray}
whereas it is maximum for the maximum value of $r_\perp$ allowed
by (\ref{criterion}), which gives
\begin{eqnarray}
A(r_\perp,\sigma_*) \simeq \frac{2 C_0}{\pi} \frac{\sigma_*}{\sqrt{1-\sigma_*^2}} \simeq 1.53 \frac{\sigma_*}{\sqrt{1-\sigma_*^2}} \, . \qquad \label{maxk}
\end{eqnarray}
Because the numerical prefactors in (\ref{lowk}) and (\ref{maxk})
differ by less than $10\%$, we can say that the anisotropy factor
of this statistically axisymmetric distribution of inertial waves
is given by expression (\ref{lowk}) within $10\%$ accuracy.
Because this anisotropy factor is almost independent of $k$, we
finally expect it to be approximately given by expression
(\ref{lowk}) for a realistic superposition of inertial waves with
different wave numbers $k$.

\end{document}